\documentclass[10pt,letterpaper]{article}

\usepackage{cogsci}
\usepackage{pslatex}
\usepackage{apacite}
\usepackage{amsmath}
\usepackage{siunitx}
\usepackage{color} 
\usepackage{graphicx}
\usepackage{subfig}
\usepackage{multicol}
\usepackage{multirow}
\usepackage{color}
\usepackage{epstopdf}

\title{Modeling the Contribution of Central Versus Peripheral Vision in Scene, Object, and Face Recognition}
 
\author{{\large \bf Panqu Wang (pawang@ucsd.edu)} \\
  Department of Electrical and Engineering,  University of California San Diego \\
  9500 Gilman Dr 0407, La Jolla, CA 92093 USA
  \AND {\large \bf Garrison W. Cottrell (gary@ucsd.edu)} \\
  Department of Computer Science and Engineering,  University of California San Diego \\
  9500 Gilman Dr 0404, La Jolla, CA 92093 USA}

\begin{document}

\maketitle

\begin{abstract}
It is commonly believed that the central visual field (fovea and parafovea) is important for recognizing objects and faces, and the peripheral region is useful for scene recognition. However, the relative importance of central versus peripheral information for object, scene, and face recognition is unclear. \citeA{larson2009contributions} investigated this question in the context of scene processing using experimental conditions where a circular region only reveals the central visual field and blocks peripheral information ("Window"), and in a "Scotoma" condition, where only the peripheral region is available. They measured the scene recognition accuracy as a function of visual angle, and demonstrated that peripheral vision was indeed more useful in recognizing scenes than central vision in terms of achieving maximum recognition accuracy. In this work, we modeled and replicated the result of \citeA{larson2009contributions}, using deep convolutional neural networks (CNNs). Having fit the data for scenes, we used the model to predict future data for large-scale scene recognition as well as for objects and faces. Our results suggest that the relative order of importance of using central visual field information is face recognition$>$object recognition$>$scene recognition, and vice-versa for peripheral information. Furthermore, our results predict that central information is more efficient than peripheral information on a per-pixel basis across all categories, which is consistent with Larson and Loschky's data.

\end{abstract}

\section{Introduction}
Viewing a real-world scene occupies the entire visual field, but the visual resolution across the visual field varies. The fovea, a small region in the center of the visual field that subtends approximately $\SI{1}{\degree}$  of visual angle \cite{polyak1941retina}, perceives the highest visual resolution of $20$ to $45$ cycles/degree (cpd) \cite{loschky2005limits}. The parafovea has a slightly lower visual resolution and extends to about $4$-$\SI{5}{\degree}$ eccentricity, where the highest density of rods is found \cite{wandell1995foundations}. Beyond the parafovea is generally considered to be peripheral vision \cite{holmes1977peripheral}, which receives the lowest visual resolution. Due to the high density and small receptive field of retinal receptors,the central (foveal and parafoveal) vision encodes information of higher spatial frequency and more detail; peripheral vision, on the contrary, encodes coarser and lower spatial frequency information.

This retinotopic representation of the visual field is mapped to visual cortical areas through a log-polar representation. Recent studies have shown that orderly central and peripheral representations can be found not only in low-level to mid-level visual areas (V1-V4), but also in higher-level regions, where perception and recognition for faces or scenes is engaged \cite{malach2002topography,grill2004human}. More specifically, \citeA{malach2002topography} proposed that the need for visual resolution is a crucial factor in organizing object areas in higher-level visual cortex: object recognition that depends more on fine detail is associated with central-biased representations, such as faces and words; object recognition that depends more on large-scale integration is associated with peripheral-biased representations, such as buildings and scenes. This hypothesis is supported by fMRI evidence, which shows that the brain areas that are more activated for faces (FFA; \citeA{kanwisher1997fusiform}) and words (VWFA; \citeA{mccandliss2003visual}) sit in the eccentricity band expanded by central visual-field bias, whereas buildings and scenes (PPA; \citeA{epstein1999parahippocampal}) are associated with peripheral bias. More recent studies even suggest that the central-biased pathway for recognizing faces and peripheral-biased pathway for recognizing scenes are segregated by mid-fusiform sulcus (MFS) to enable fast parallel processing \cite{gomez2015functionally}.

In the domain of behavioral research, studies have shown that object perception performance is the best around $\SI{1}{\degree}$-$\SI{2}{\degree}$ of fixation point and drops rapidly as eccentricity increases \cite{henderson1999role,nelson1980functional}. For scene recognition, \citeA{larson2009contributions} used a  "Window" and "Scotoma" design (see Figure \ref{CogSci2016Fig1-01}), to test the contributions of central versus peripheral vision to scene recognition. The Window condition (top rows of the right-hand columns of Figure \ref{CogSci2016Fig1-01}) presents central information at various visual angles to the subjects, while the Scotoma condition (second row on the right) blocks it. Using images from 10 categories, subjects were required to verify the category in each condition. The recognition accuracy as a function of visual angle is shown in Figure \ref{CogSci2016Fig2-01}. They found that foveal vision is not accurate for scene perception, while peripheral vision is, despite its much lower resolution. However, they also found that central vision is more efficient, in the sense that less area is needed to achieve equal accuracy. The visual area is equal at $\SI{10.8}{\degree}$, and the crossover point, where central vision starts to perform better than peripheral, is to the left of that point. 

Despite the common belief that central vision is important for face and object recognition, and peripheral vision is important for scene perception shown in studies above, a careful examination of the contribution of central versus peripheral vision in object, scene, and face recognition is needed. In this work, we modeled the experiment of \citeA{larson2009contributions} using deep convolutional neural networks. Furthermore, we extended the modeling work to a greater range of stimuli, and answer the following questions: How does the model perform as the number of scene categories is scaled up? Besides scenes, can the model predict the importance of central vision versus peripheral information in object and face recognition? What is the result compared to scenes?

In the following, we show that our modeling results match the observations of \citeA{larson2009contributions}, and that it scales up to over 200 scene categories. By running a similar analysis for large-scale object and face recognition, our model predicts that central vision is very important for face recognition, important for object recognition, and less important for scene recognition. Peripheral vision, however, serves an important role for scene recognition, but is less important for recognizing objects and faces. Furthermore, across all conditions we tried, central vision is more efficient than peripheral vision on a per-pixel basis (when equal areas are presented), which is consistent with the result of \citeA{larson2009contributions}. 

\section{Method}
\subsection{Image Preprocessing}
\begin{figure}[t]
\begin{center}
\includegraphics[width=0.5\textwidth]{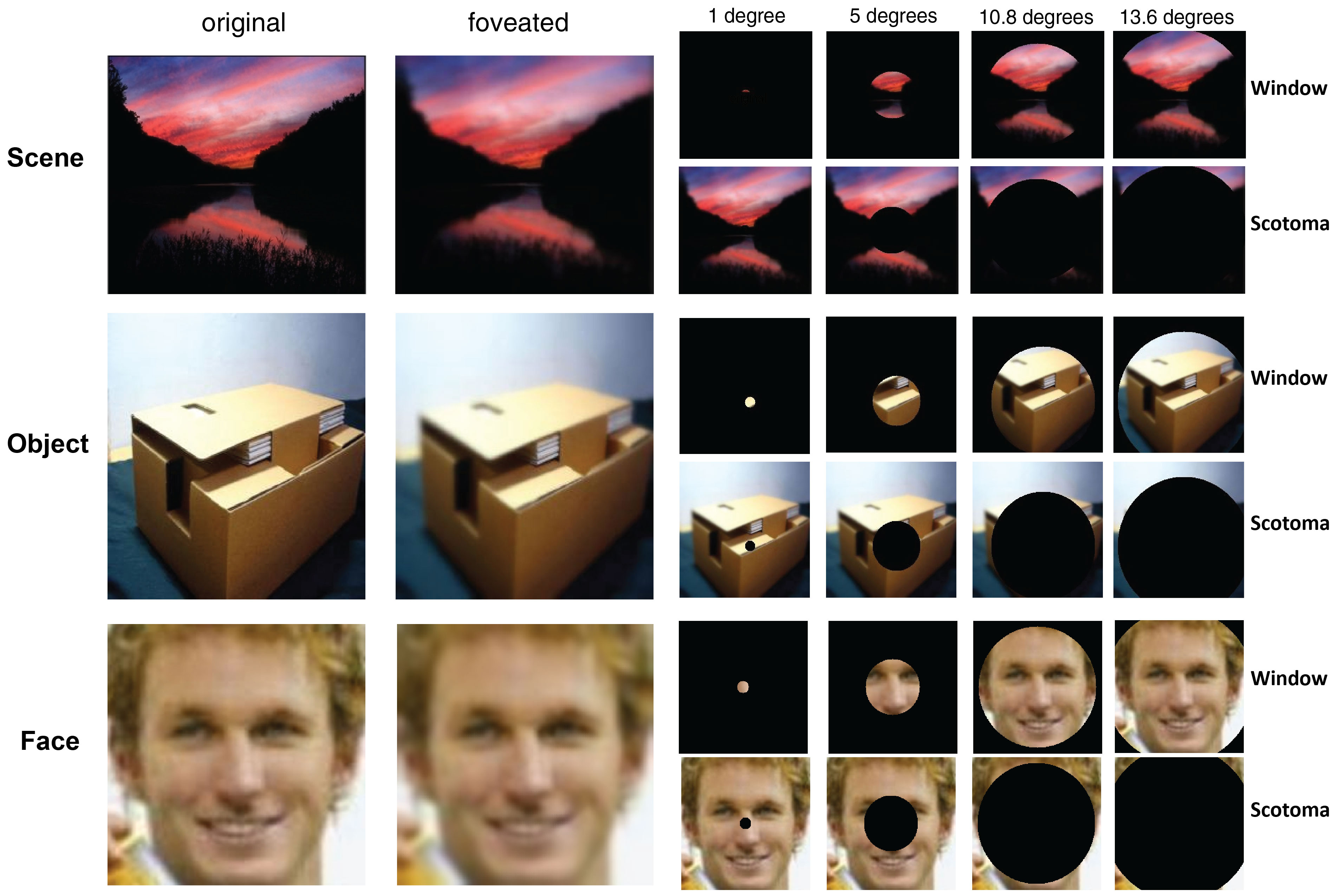}
\end{center}
\caption{Examples of images used in our experiment. First column: original images. Second column: foveated images. Third to last column: images processed through "Window" and "Scotoma" conditions with different radii in degrees of visual angle.}
\vspace{-10pt}
\label{CogSci2016Fig1-01}
\end{figure}

To create foveated images, we preprocessed the images using the Space Variant Imaging System\footnote{\url{http://svi.cps.utexas.edu/software.shtml}}. To mimic human vision, we set the parameter that specifies the eccentricity at which resolution drops to half of the fovea to $\SI{2.3}{\degree}$. Example images and their preprocessed retinal versions are shown in the first and second columns of Figure~\ref{CogSci2016Fig1-01}.

As in the experiments of \citeA{larson2009contributions}, we used the Window and Scotoma paradigms as specified by \citeA{van1998functional} to process the input stimulus. The idea of both paradigms is to evaluate the value of missing information - if the missing information is needed, then the perception process may be disrupted and recognition performance may drop; if the missing information is not necessary, then the processing remains normal. 

Input images in our experiments are  $256\times256$ pixels, and we assume that corresponds to $\SI{27}{\degree}\times\SI{27}{\degree}$ of visual angle, the number in \cite{larson2009contributions}. In \cite{larson2009contributions}, they used four sets of radius conditions for Windows and Scotomas: $\SI{1}{\degree}$ represents the presence or absence of foveal vision; $\SI{5}{\degree}$ represents the presence or absence of central vision; $\SI{10.8}{\degree}$ presents equal viewable area inside the Windows or outside the Scotomas; $\SI{13.6}{\degree}$ presents more viewable area in the Windows than the Scotomas. In order make the prediction of the model more accurate, we added five additional radius conditions in all of our experiments: $\SI{3}{\degree},\SI{7}{\degree},\SI{9}{\degree},\SI{12}{\degree}$, and $\SI{16}{\degree}$. The example Window and Scotoma images are shown in Figure~\ref{CogSci2016Fig1-01}.

\subsection{Deep Convolutional Neural Networks (CNNs)}
Deep CNNs are neural networks with many layers that stack computations in a hierarchical way, repeatedly performing: 1) 2-dimensional convolutions over the stimulus generated from previous layers using learned filters, which are connected locally to a small subregion of the visual field; 2) a pooling operation on local regions of the feature maps obtained from convolution operation, which is used to reduce the dimensionality and gain translational invariance; 3) non-linearities to the upstream response, which is used to generate more discriminative features useful for the task. As layers go higher, the receptive fields of filters are generally larger, and the learned features go from low-level (edges, contours) to high-level object-related representations (object parts and shapes) \cite{zeiler2014visualizing}. Several fully-connected layers are usually added on top of these computations to learn more abstract and task-related features.

We used deep CNNs in our experiments for two reasons. First, deep CNNs are the best models in computer vision: they achieve the state-of-the-art performance on many large-scale computer vision tasks, such as image classification \cite{krizhevsky2012,he2015deep}, object detection \cite{ren2015faster}, and scene recognition \cite{zhou2014learning}. Thus, the models should achieve decent performance in our experiments. Smaller networks or other algorithms are not competent for our tasks. Second, deep CNNs have been shown to be the best models of the visual cortex: they are able to explain a variety of neural data in human and monkey IT \cite{yamins2014performance,gucclu2015deep,wang2015encoding}. As a result, it is natural to use them in our work modeling a behavioral study related to human vision.

\section{Experiments}
In this section, we first describe our model of the behavioral study of \citeA{larson2009contributions}. We then introduce the experiment for measuring the contribution of central versus peripheral vision for large-scale scene, object, and face recognition tasks.
\subsection{Modeling \citeA{larson2009contributions}}
\begin{figure}[t]
\begin{center}
\includegraphics[width=0.5\textwidth]{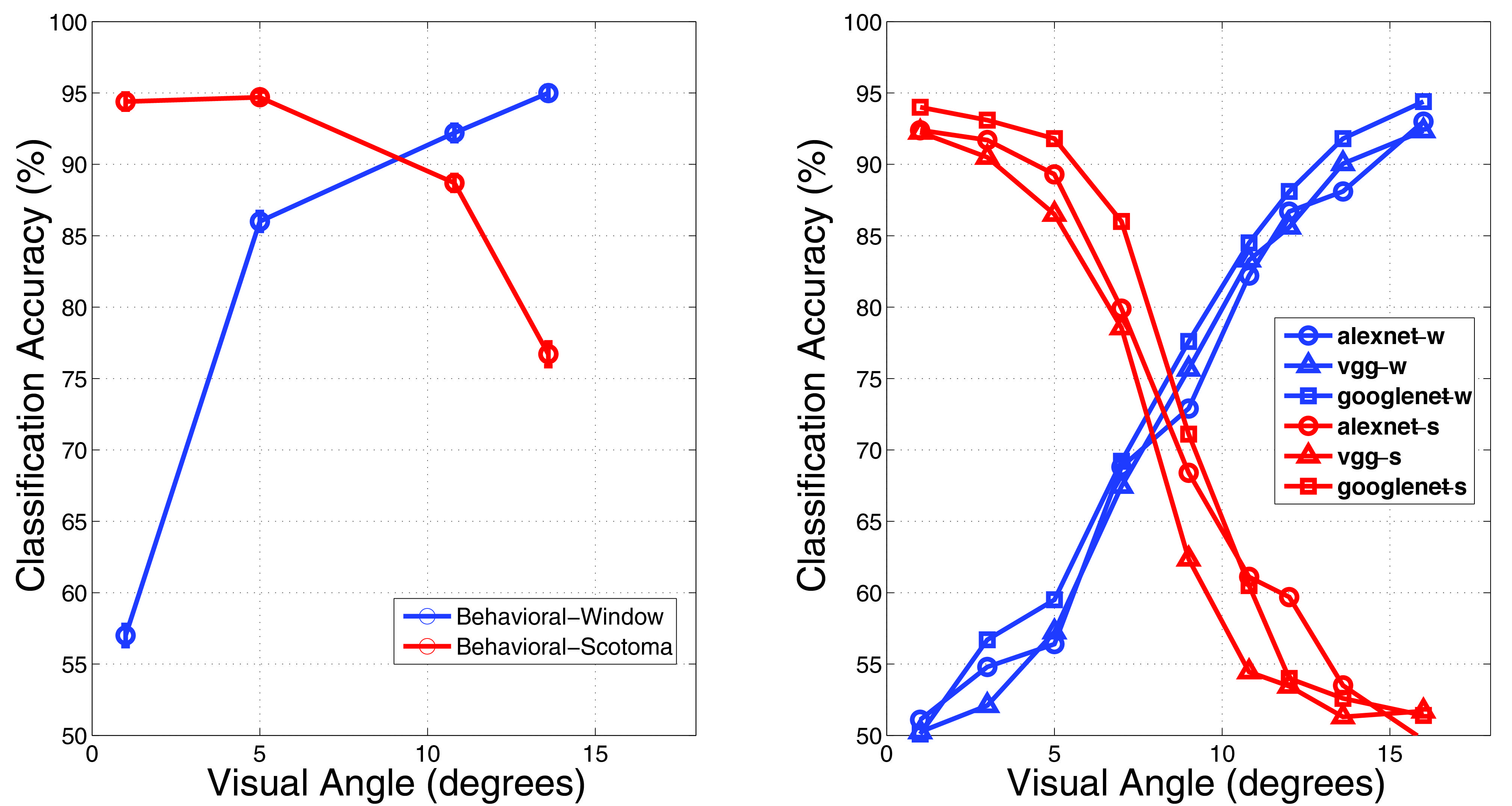}
\end{center}
\caption{Results for scene recognition accuracy as a function of viewing condition (Windows (w) and Scotomas (s)) and visual angle. Left: result of \citeA{larson2009contributions}. Right: our modeling result. }
\vspace{-0.3in}
\label{CogSci2016Fig2-01}
\end{figure}

In \citeA{larson2009contributions}, scene recognition accuracy was measured across 100 human subjects on 10 categories: Beach, Desert, Forest, Mountain, River, Farm, Home, Market, Pool, and Street. For each trial in the Windows and Scotomas conditions, subjects were first presented a scene image, and then were asked to press "yes" or "no" for the cue (category name) presented on the screen. Their experimental result is summarized in Figure~\ref{CogSci2016Fig2-01}. They showed that central vision ($\SI{5}{\degree}$ window condition) performs less well than peripheral vision in terms of getting maximum recognition performance. They further demonstrated the peripheral advantage is due to more viewing areas in the Scotomas conditions, and central vision is more privileged when given equal viewable areas ($\SI{10.8}{\degree}$).

We obtained the stimuli of the above 10 categories from the Places205 database \cite{zhou2014learning}, which contains 205 scene categories and 2.5 million images. All input stimuli were preprocessed using the retina model described in the above section. As 10 categories is small and can easily lead to overfitting problems in training deep CNNs, we trained our recognition model by performing fine-tuning (or transfer learning) based on pretrained models. The model pretrained on the Places205 database can be treated as a mature scene recognition pathway, and fine-tuning can be thought as additional training for the task. To investigate whether different network architectures, especially depth, have different impact on the modeling result, we applied three different pre-trained models, namely:
\begin{enumerate}
  \item AlexNet \cite{krizhevsky2012}: A network with 5 convolutional layers and 3 fully connected layers, about 60 million trainable parameters. Achieved $81.10\%$ top-5 accuracy on the Places205 validation set. 
  \item VGG-16 \cite{simonyan2014very}: A network with 13 convolutional layers and 3 fully connected layers, about 138 million trainable parameters. Achieved $85.41\%$ top-5 accuracy on the Places205 validation set. 
  \item GoogLeNet \cite{Szegedy_2015_CVPR}: A network with 21 convolutional layers and 1 fully connected layer, about 6.8 million trainable parameters. Achieved $87.70\%$ top-5 accuracy on the Places205 validation set. 
\end{enumerate} 
For all models, the fine-tuning process starts by keeping the weights except for the last fully connected layer intact, and initializing the weights of the last layer to be random with zero mean and unit variance. To be compatible with the "yes" or "no" condition in the behavioral experiment, we replaced the last layer in the networks with a single logistic unit, and trained the networks for each of the 10 object categories separately, using half of the training images from the target category and half randomly selected from all other 9 categories. As the last layer needs more learning, we set the learning rate of the last layer to 0.001, and all previous layers to $1e^{-4}$. The training set of the 10 scene categories contains a total number of 129,210 full resolution images, and we trained all networks using minibatch stochastic gradient descent with batch size from 32 to 256, using the Caffe deep learning framework \cite{jia2014caffe} on NVIDIA Titan Black 6GB GPUs. All networks were trained for a maximum number of 24,000 iterations to ensure convergence. Each test set contains 200 images (100 from target category and 100 from all other categories), and the label distribution is the same as the training set.. Test images were preprocessed to meet each of the Windows and Scotomas condition. We tested the performance of the fine-tuned models on all conditions by reporting the mean classification accuracy, which is shown in Figure~\ref{CogSci2016Fig2-01}.

From Figure~\ref{CogSci2016Fig2-01}, we can clearly see our result for all three models qualitatively matches the result of \citeA{larson2009contributions}. First, for Window and Scotoma conditions, an increasing radius of visual angle (x axis) yields a monotonic increase or decrease in classification accuracy (y axis). The sharper increase from $\SI{1}{\degree}$ to $\SI{5}{\degree}$ in the behavioral study may be due to the higher efficiency of human central vision. Second, we replicated the fact that central vision (less than $\SI{5}{\degree}$) is less useful than peripheral vision in terms achieving the best scene recognition performance. Third, however, when using equal viewable areas ($\SI{10.8}{\degree}$), central vision performs better than peripheral, exhibiting higher efficiency. Fourth, the critical radius (the crossover point where the two conditions produce equal performance, see Figure~\ref{CogSci2016Fig2-01}b) is $\SI{8.26}{\degree}$ (averaged across all models), which is within the $\SI{8.22}{\degree}$-$\SI{9.24}{\degree}$ range reported by \citeA{larson2009contributions}.
This suggests our models are quite plausible.

When comparing the performance across the three models we use, we cannot find a notable difference in terms of performance, though GoogLeNet usually performs slightly better, indicating that depth of processing might be the key factor in obtaining better performance. 
\subsection{Large-Scale Scene, Object, and Face Recognition}
The above modeling work is based on a scene recognition task using 10 categories. In real life, however, there are a much larger number of scene categories. Beyond scenes, general object recognition and face recognition are the two most important recognition tasks that are performed regularly. The relative importance of central versus peripheral vision among the three categories needs to be examined carefully. Using a similar modeling approach, we describe our findings in large-scale scene, object, and face recognition in the sections below.
\subsubsection{Scene Recognition}
\begin{figure}[t]
\begin{center}
\includegraphics[width=0.5\textwidth]{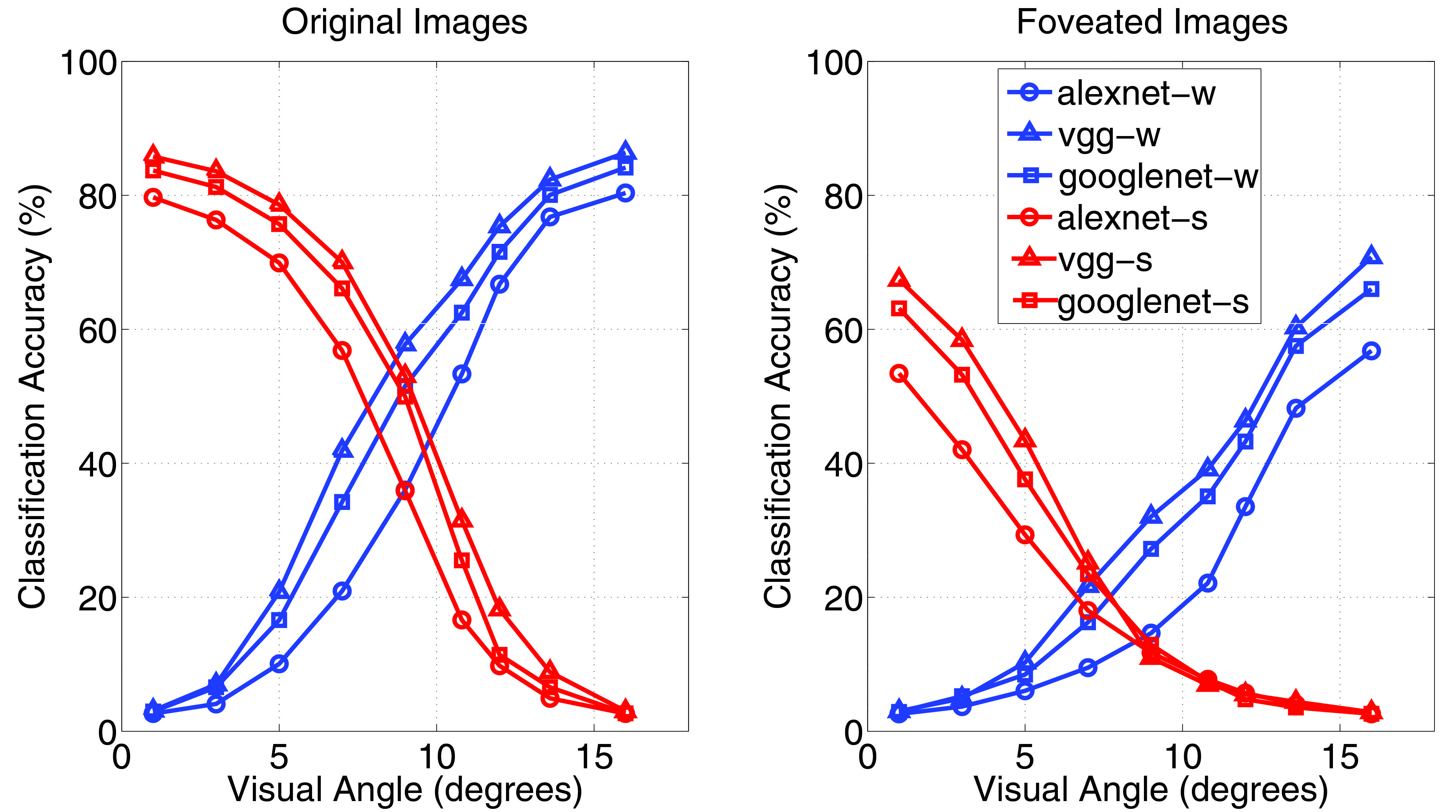}
\end{center}
\caption{Results for large-scale scene recognition accuracy as a function of viewing condition (Windows (w) and Scotomas (s)) and visual angle. Softmax output is used instead of logistic unit, so chance is 0.005. Left: experiment using original images. Right: experiment using foveated images. }
\vspace{-0.2in}
\label{CogSci2016Fig3-01}
\end{figure}

We used all 205 categories in the Places205 dataset. The trained models of AlexNet, VGG-16, and GoogLeNet are deployed to examine the recognition accuracy on the Place205 validation set, which contains $20,500$ images, in all Windows and Scotoma conditions. In addition, we tested the models using images both processed and unprocessed by the retina model to examine the generalization power of the learned features. The result is shown in Figure~\ref{CogSci2016Fig3-01}.
\vspace{-0.175in}

From Figure~\ref{CogSci2016Fig3-01}, we can see the general trend that we observed in Figure~\ref{CogSci2016Fig2-01} still holds: peripheral vision is more important than central vision, but central vision is more efficient. All models behave similarly. However, we can see the performance on images preprocessed through the retina model is inferior. Apparently, since there are many more categories in this experiment, the foveation has more of an effect. Recall that the models are trained using images with full resolution; missing the peripheral information may the cause learned features to imperfectly generalize. 
\subsubsection{Object Recognition}
\begin{figure}[t]
\begin{center}
\includegraphics[width=0.5\textwidth]{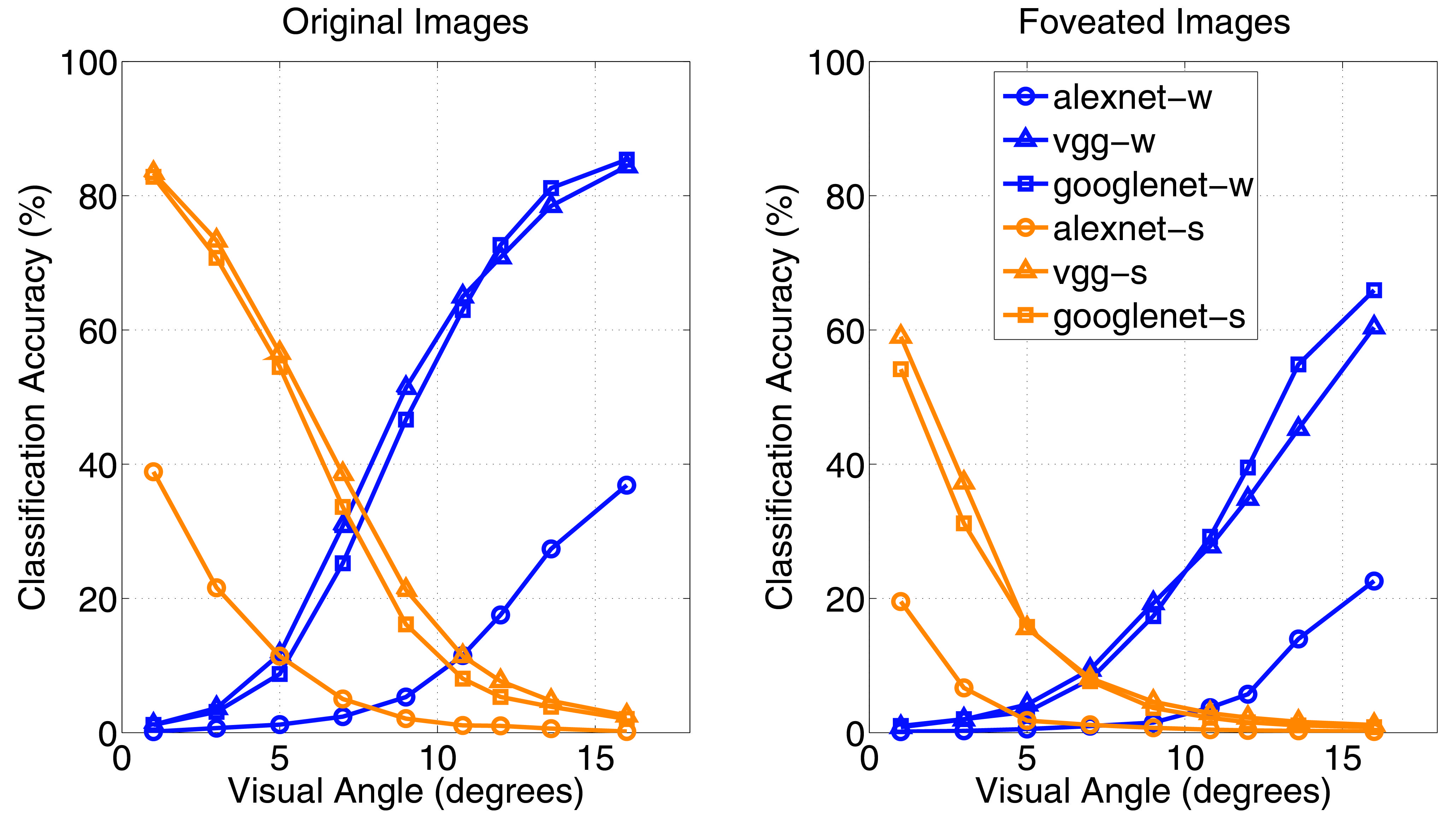}
\end{center}
\caption{Results for large-scale object recognition accuracy as a function of viewing condition (Windows (w) and Scotomas (s)) and visual angle. Softmax output is used instead of logistic unit, so chance is 0.001. Left: experiment with original images. Right: experiment with foveated images.}
\label{CogSci2016Fig4-01}
\vspace{-0.2in}
\end{figure}

We ran our object recognition experiment on the ILSVRC 2012 dataset \cite{ImageNetDataset}, which contains 1000 object categories and over 1.2 million training images. We used the pretrained models of AlexNet, VGG-16, and GoogLeNet, which achieve top-5 accuracy of $80.13\%$, $88.44\%$, and $89.00\%$, respectively, on the ILSVRC 2012 validation set. Similar to scene recognition, we tested all models under all Windows and Scotoma conditions, using original and foveated images. The results are shown in Figure~\ref{CogSci2016Fig4-01}.

\begin{figure}[h]
\begin{center}
\includegraphics[width=0.5\textwidth]{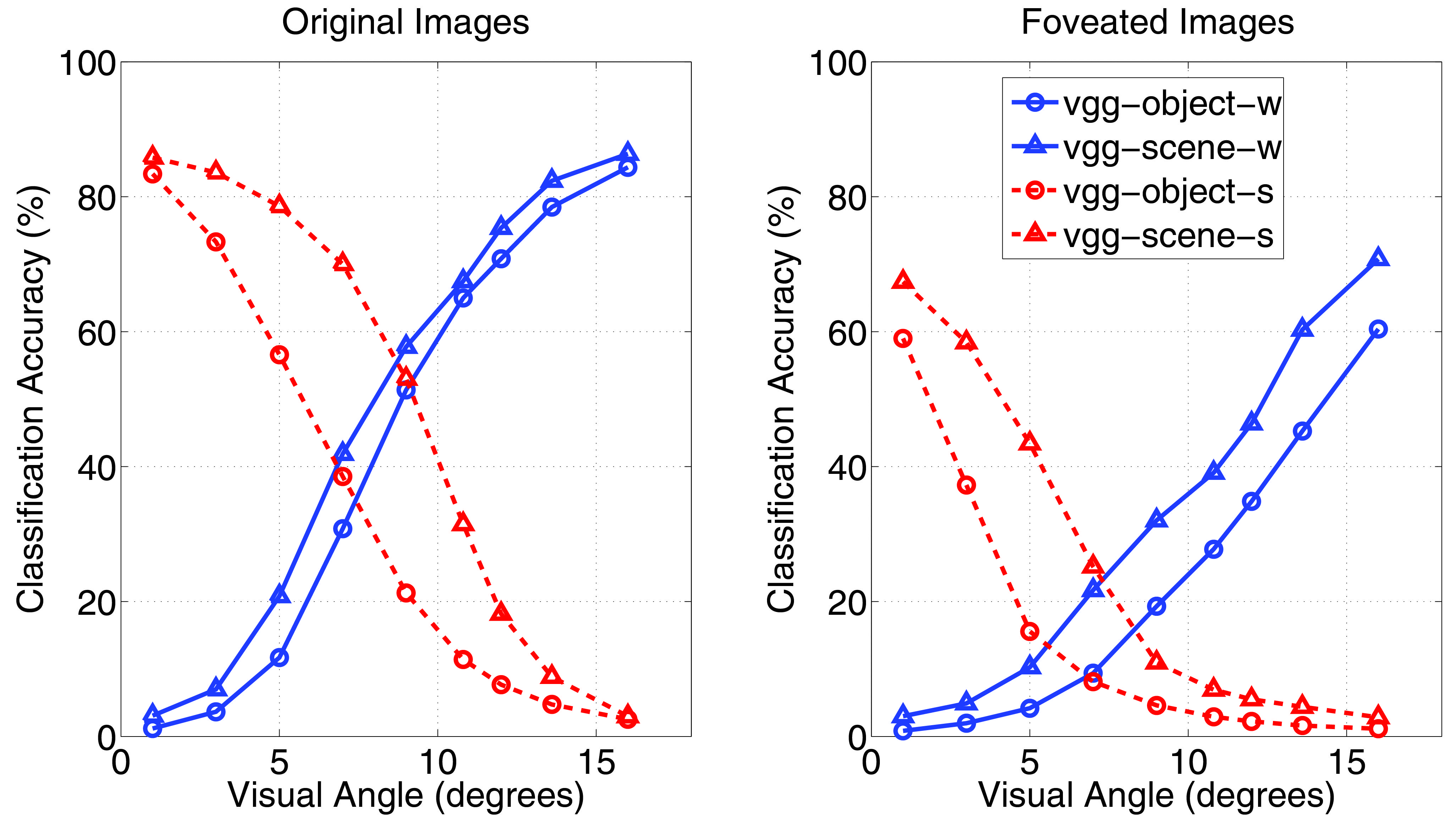}
\end{center}
\caption{Comparison results for scene and object recognition using the VGG-16 model. Losing central vision decreases performance for object recognition more quickly than scene recognition. Left: original images. Right: foveated images. }
\label{CogSci2016Fig5-01}
\vspace{-0.2in}
\end{figure}

At the first glance of looking at Figure~\ref{CogSci2016Fig4-01}, we may draw the conclusion that the result is the same as scene recognition: central vision is still more important than peripheral vision. However, when we compare the scene and object recognition results (shown in Figure~\ref{CogSci2016Fig5-01}), we can clearly see that central information in object recognition is more important than that in scene recognition: the accuracy of the Scotoma conditions drops much faster for object recognition than scene recognition as visual angle increases from $\SI{1}{\degree}$ to $\SI{7}{\degree}$, suggesting that losing  central vision causes a greater impairment for object recognition performance. This is consistent with our knowledge that central vision plays a more important role in object recognition than scenes, as there are more high spatial frequency details in objects than scenes. Another finding from this experiment is that AlexNet (8 layers) performs much worse than VGG-16 (16 layers) and GoogLeNet (23 layers), suggesting that depth is important to produce good performance. 
\vspace{-0.0in}
\subsubsection{Face Recognition}
\begin{figure}[t]
\begin{center}
\includegraphics[width=0.5\textwidth]{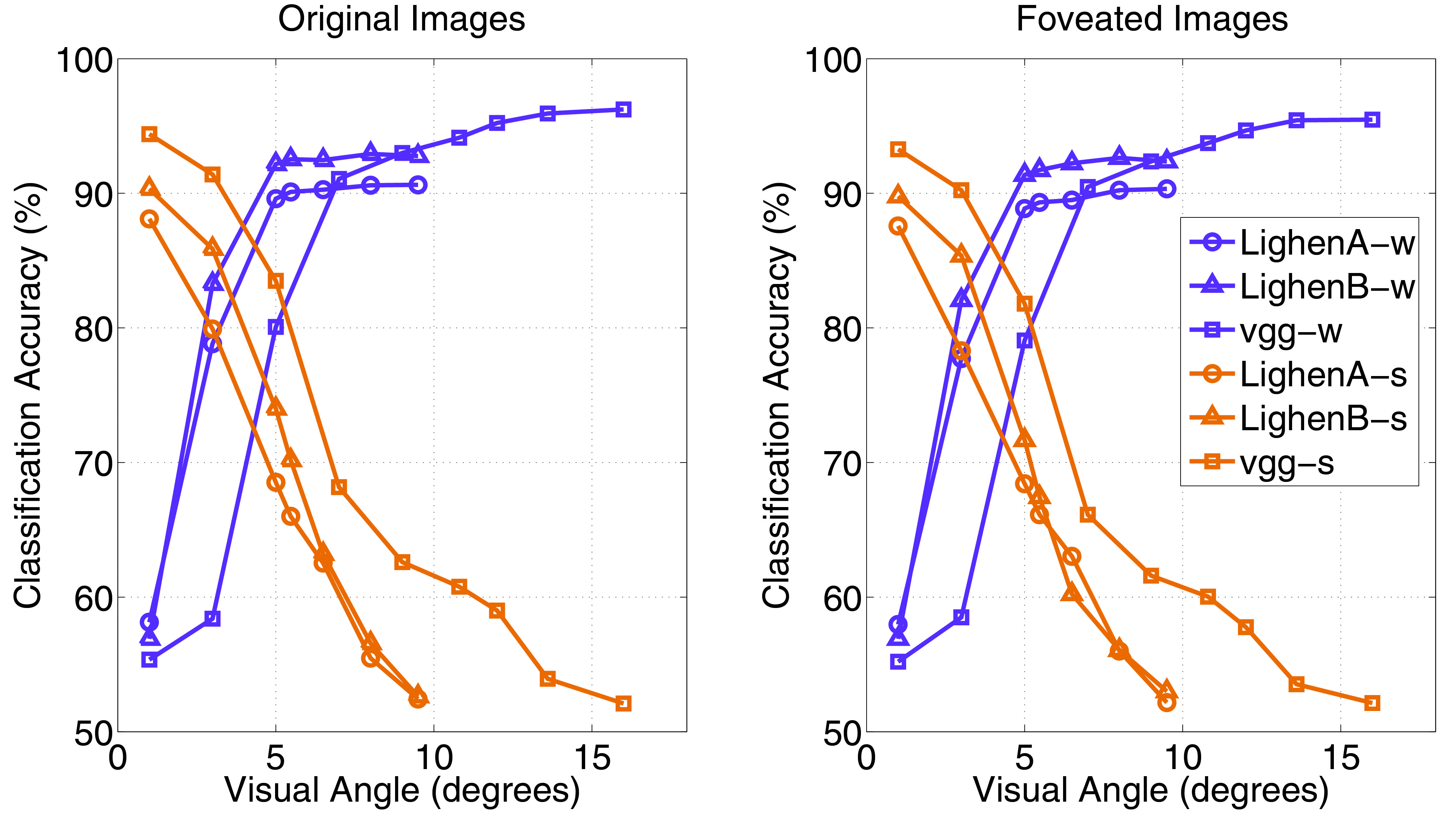}
\end{center}
\caption{Results for large-scale face recognition accuracy as a function of viewing condition (Windows (w) and Scotomas (s)) and visual angle. Left: experiment with original images. Right: experiment with foveated images. For Lighten-A and Lighten-B models, the visual angle only expands to $\SI{9.5}{\degree}$, as the input image is smaller ($144\times144$) than for the VGG model ($256\times256$). The accuracy for face verification task is measured as the true positive rate at Equal Error Rate (EER) point on the ROC curve. Chance is 0.5.}
\vspace{-10pt}
\label{CogSci2016Fig6-01}
\end{figure}

We performed the face recognition experiment on the Labeled Faces in the Wild (LFW) dataset \cite{LFWTech}, which contains $13,233$ labeled images from $5,749$ individuals. As there is only 1 image for some identities, researchers usually pretrain their network on larger datasets (not publicly available) and test their models on the LFW dataset. In this experiment, we tested three pretrained models, namely Lighten-A (10 layers; \cite{wu2015lightened}), Lighten-B (16 layers), and VGG-Face (16 layers;\cite{Parkhi15}), on the face verification task for the LFW dataset, where they achieve accuracy of $90.33\%$, $92.37\%$, and $96.23\%$, respectively. Face images were preprocessed so that they occupy the entire visual field (Figure~\ref{CogSci2016Fig1-01}). Same as the previous experiments, we tested all models using Windows and Scotoma conditions, with original and foveated images. Results are shown in Figure~\ref{CogSci2016Fig6-01}.

We see very different performance in  Figure~\ref{CogSci2016Fig6-01} compared to object and scene recognition. First, central information is obviously much more important than peripheral information for face recognition, given the accuracy at $\SI{5}{\degree}$ is much higher for the Window condition than the Scotoma condition for Lighten models, and very similar with each other for the VGG model. This is consistent with our intuition that face recognition is a fine-grained discrimination process. Second, the Window performance grows much more slowly after $\SI{7}{\degree}$, suggesting the more peripheral region provides little additional information for recognizing faces, unlike objects and scenes, which needs lots of peripheral information to obtain the maximal accuracy. Third, the foveated images produce nearly identical results as the original image, demonstrating that face recognition only involves central vision, and the blurred peripheral vision is not needed. 

\begin{figure}[t]
\begin{center}
\includegraphics[width=0.5\textwidth]{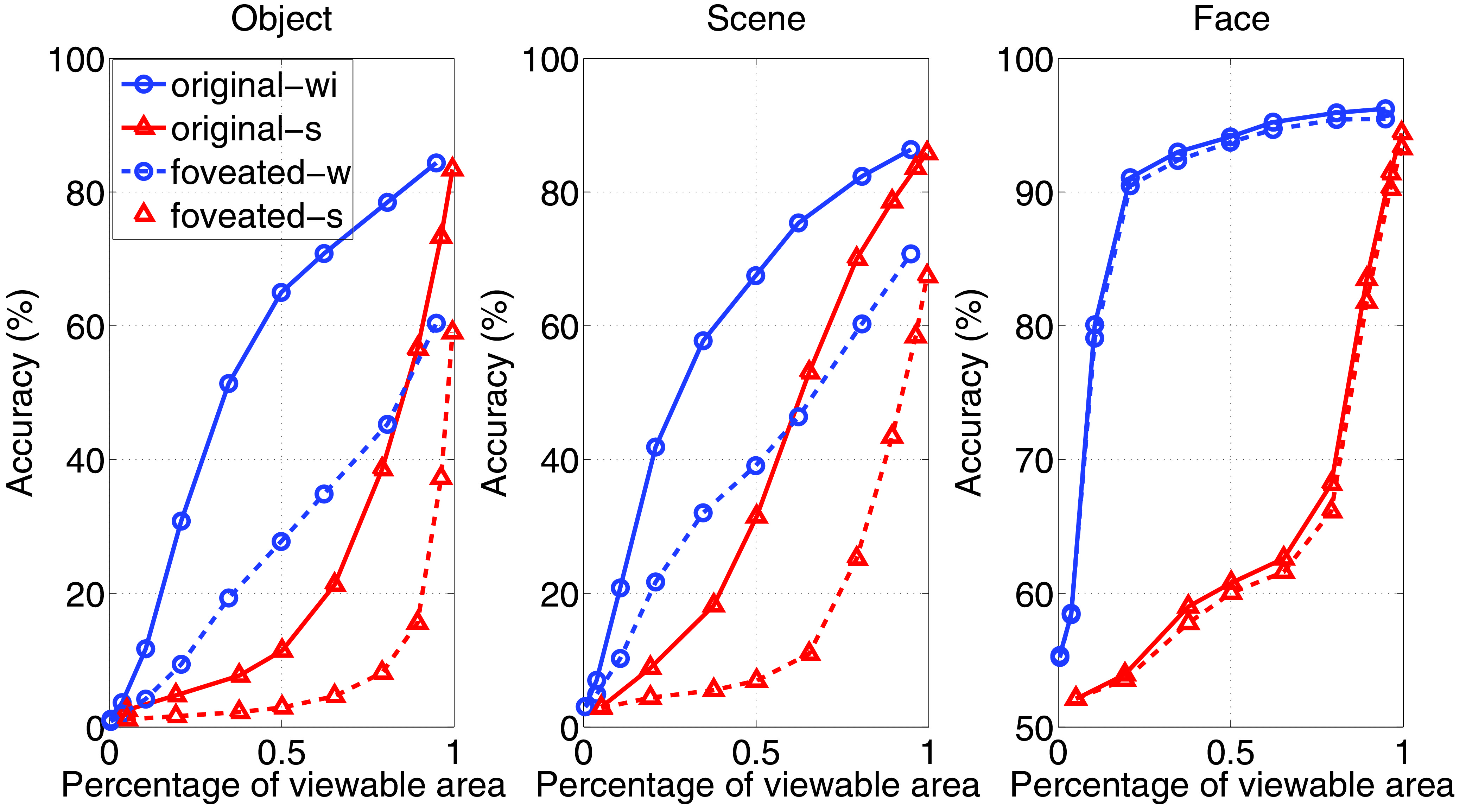}
\end{center}
\caption{Accuracy for object (left), scene (middle) and face (right) recognition as a function of the percentage of viewable area presented under Window (blue) and Scotoma (red) conditions, using original (solid line) and foveated images (dashed line).}
\vspace{-10pt}
\label{CogSci2016Fig7-01}
\end{figure}

Finally, as central vision appears to be more efficient (on a per-pixel basis) than peripheral vision in all experiments we tried, we tested the relative efficiency of the central vision over peripheral vision by measuring the recognition accuracy as a function of viewable area. The result is shown in Figure~\ref{CogSci2016Fig7-01}.

From Figure~\ref{CogSci2016Fig7-01}, we can clearly see that the recognition accuracy of central vision is always superior than peripheral vision for all tasks. However, central vision is even more efficient when recognizing faces than recognizing objects or scenes, as viewable areas over $50\%$ of the whole image can only provide a limited boost for face recognition, while significantly improving the accuracy of object and scene recognition. Contrarywise, peripheral information provides little to no help for face recognition, unless over $90\%$ of the image is presented, but the accuracy still suffers due to the loss of central vision. However, peripheral information is important for object and scene recognition (and more important for scene recognition, as shown in Figure~\ref{CogSci2016Fig5-01}).

These large-scale scene, object, and face recognition modeling results suggest there is an order of relative importance of central versus peripheral vision in those tasks: peripheral vision is most important for scene recognition, less important for object recognition, and basically not helpful for face recognition. Central vision, however, plays a crucial role in face recognition, is important for object recognition, and is less important for scene recognition.

\section{Conclusion}
In this paper, we modeled the contribution of central versus peripheral visual information for scene, object, and face recognition, using deep CNNs. We first modeled the behavioral study of \citeA{larson2009contributions}, and replicated their findings of the importance of peripheral vision in scene recognition. In addition, by running a large-scale scene, object, and face recognition simulation, our models make testable predictions for the relative order of importance for central versus peripheral vision for those tasks. 


\section{Acknowledgments}
This work was supported by NSF grants IIS-1219252 and SMA 1041755 to GWC. PW was supported by a fellowship from Hewlett-Packard.

\bibliographystyle{apacite}

\setlength{\bibleftmargin}{.125in}
\setlength{\bibindent}{-\bibleftmargin}

\bibliography{CogSci2016}

\end{document}